\documentclass[11pt,a4paper]{article}

\usepackage[margin=2.4cm]{geometry}
\usepackage[T1]{fontenc}
\usepackage[utf8]{inputenc}
\usepackage{lmodern}
\usepackage{hyperref}
\usepackage{setspace}
\usepackage{enumitem}
\usepackage{amsmath}
\usepackage{amssymb}
\usepackage{wrapfig}
\usepackage{graphicx}
\usepackage{caption}

\usepackage{natbib}
\newcommand{\Msun}{M_\odot}
\newcommand{\Mstar}{M_\star}

\usepackage{multicol}

\begin{document}

\begin{center}

    {\Huge \textbf{Searching for Intermediate-Mass Black Holes in Milky Way satellites}}\\[3em]
    
    {\Large R. Pascale$^1$, G. Battaglia$^2$}\\[1.5em]
    
    {\large 
    $^1$  INAF – Osservatorio di Astrofisica e Scienza dello Spazio di Bologna, Via Piero Gobetti 93/3, 40129 Bologna, Italy \\
    $^2$ Instituto de Astrofísica de Canarias, Calle Vía Láctea s/n, 38206 La Laguna, Santa Cruz de Tenerife, Spain\\
    }
\end{center}
\clearpage

\section*{Summary}

Intermediate-mass black holes (IMBHs), with masses between roughly $10^2\Msun$ and $10^5\Msun$ \citep{Krajnovic2018,Greene2020}, represent a largely uncharted component of the black-hole (BH) population. They are theoretically predicted to form in several early-Universe pathways, including the remnants of massive Population III stars, the runaway collapse of dense stellar clusters, and the direct collapse of metal-poor gas \citep{Madau2001,Volonteri2003,Somerville2008,Greene2020}. Establishing whether IMBHs are present in dwarf galaxy satellites of the Milky Way (MW), and with what occupation fraction - i.e. the fraction of galaxies with a certain stellar mass that host a central BH - provides one of the most incisive tests of BH seed formation models.

Despite their importance, present dynamical constraints on IMBHs remain weak. Dynamical IMBH masses or upper limits are available for very few such systems \citep{Lora2009,Jardel2012,Pascale2024}, with secure detections in less than ten cases \citep[e.g.][]{denBrok,Nguyen2018,Nguyen2019ApJ}. A next-generation wide-field spectroscopic facility, capable of combining deep multiplexed stellar spectroscopy with high-resolution integral-field observations of galaxy centers, would open access to IMBH masses in the $\leq10^4 \Msun$ regime. Such an advance would make possible - for the first time - a robust measurement of the IMBH occupation fraction in dwarf galaxies.

{\bf A key scientific requirement for the coming decades is to establish the observational and instrumental capabilities needed to detect or tightly constrain IMBHs in nearby dwarf galaxies, particularly in the $\simeq10^3-10^4 \Msun$ mass range, and thereby enable a measurement of their occupation fraction. Such a measurement is fundamental for distinguishing between competing scenarios for the formation of BH seeds in the early Universe.}

\section*{Scientific Context}

The BH mass spectrum is well established at its extremes: stellar-mass BHs with masses $\lesssim10^2\Msun$, and supermassive black holes (SMBHs) with masses $\gtrsim
10^5-10^6 \Msun$ residing in galaxy nuclei \citep{Kormendy2013}. The intermediate regime, however, remains sparsely explored. Yet IMBHs are a natural expectation of several formation channels, and their existence may be intimately connected to the emergence of billion-solar-mass SMBHs already observed at $z>7$ \citep[e.g.;][]{Banados2018}. The early presence of such quasars strongly suggests either extremely rapid BH growth or the formation of comparatively massive seeds.

Within the $\Lambda$ Cold Dark Matter ($\Lambda$CMD) cosmological paradigm, dwarf galaxies provide a uniquely favourable environment for studying IMBHs \citep{Natarajan2021}. Their evolution is characterized by limited merger histories, shallow gravitational potentials, and inefficient gas retention, all of which reduce subsequent BH growth. Consequently, dwarf galaxies are predicted to preserve the imprint of their original BH seed populations more faithfully than massive galaxies. Theoretical models consistently indicate that the occupation fraction of central BHs should decline steeply below a stellar mass of $\Mstar\simeq10^9\Msun$ \citep{Volonteri2008,Greene2020}. This aspect makes the stellar-mass range of classical dwarf spheroidals (dSphs; $\Mstar\leq10^8\Msun$; \citealt{McConnachie2012}) and ultra-faint dwarfs  particularly diagnostic, especially combined with their relatively close distances.

\begin{wrapfigure}{r}{0.60\hsize}
\vspace{-0.5cm} 
  \centering
  \includegraphics[width=1\hsize]{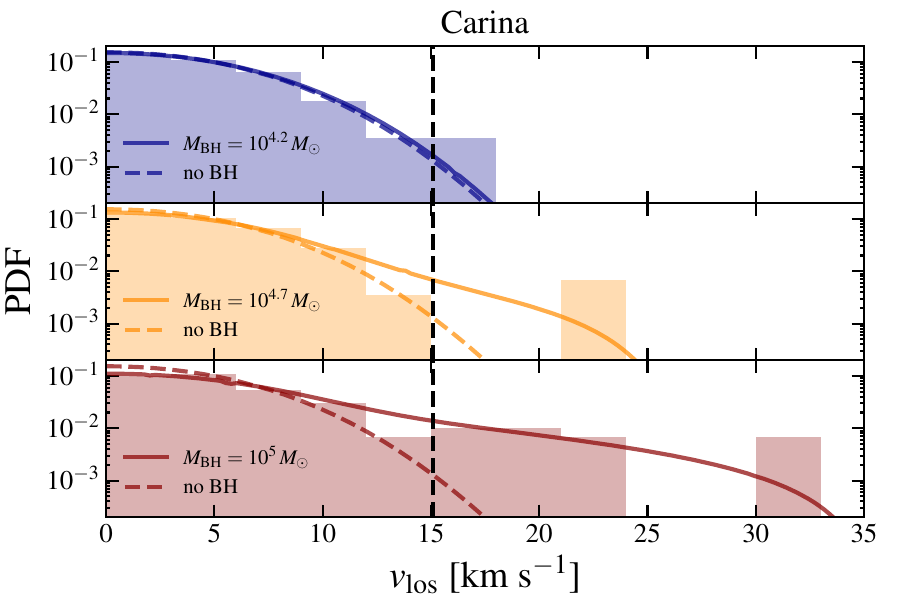}
  \captionsetup{font=small}
  \caption{Line-of-sight velocity distributions of Carina computed within the central 2 arseconds, with (solid lines) and without (dashed lines) a central IMBH. Different colors refer to different IMBH masses. The presence of an IMBH increases the probability of high-velocity stars. The histograms show a realization of 40 stars drawn from the corresponding distributions, highlighting the importance of sampling the high-velocity tail to detect putative IMBHs.}
  \label{fig:losvd}

\vspace{-0.5cm} 
\end{wrapfigure}

Direct dynamical detection of IMBHs in dwarf galaxies is, however, extremely challenging. The sphere of influence of a $10^4\Msun$ BH embedded in a system with a velocity dispersion of 7-12 km s$^{-1}$ is typically well below one parsec. At distances characteristic of dSphs ($\simeq80-250$ kpc), this corresponds to angular scales on the order of an arcsecond. Compounding this, the central potential of dSphs is heavily dominated by dark matter \citep{Hayashi2020}, which suppresses the expected rise in velocity dispersion near the galaxy center. As a result, the dynamical signature of an IMBH often manifests not as a clear central cusp, but rather through the presence of a small number of stars exhibiting anomalously high velocities - stars that populate the high-velocity wings of the line-of-sight velocity distribution  (see Fig~\ref{fig:losvd}). Recent work on compact stellar systems has shown that even a handful of such outliers can constitute strong evidence for an IMBH \citep{Haberle2024}.

Dwarf galaxy satellites of the MW thus offer a compelling opportunity. Their proximity ensures that the relevant angular scales are resolvable, while their old, gas-poor stellar populations produce clean kinematic tracers unaffected by dust or star formation. Several systems - including Carina, Sextans, Sculptor, Fornax, Leo I and Leo II - contain tens to hundreds of stars within a few parsecs of their photometric centers \citep{McConnachie2020,Battaglia2022}. These regions are precisely where high-precision stellar velocities can reveal the presence of IMBHs in the $10^3-10^4 \Msun$ mass range. Thus, robust detections require accurate global dynamical modeling \citep{Pascale2018,Pascale2025}, together with a sufficiently large number of high-quality kinematic measurements in the innermost arcseconds (see Fig.~\ref{fig:losvd}).

\section*{Instrumentation and Facility Requirements}
Detecting or constraining IMBHs in local dwarf galaxies requires a combination of observational capabilities that no existing facility currently provides in a single platform. The scientific challenge is intrinsically twofold. On one side, robust dynamical modeling of dwarf galaxies demands wide-field spectroscopic coverage extending to several half-light radii, in order to disentangle the respective roles of stellar anisotropy, dark matter, and any central compact object. On the other, the detection of an IMBH - particularly in the mass range $\simeq10^3-10^4 \Msun$ - depends critically on resolving and characterizing stellar motions within the extremely small sphere of influence, often confined to the central arcsecond.

A future spectroscopic facility must therefore combine deep multiplexed observations across fields of at least one square degree with high spatial-resolution spectroscopy of the central regions. Achieving velocity uncertainties of $\leq2$ km s$^{-1}$ for stars down to magnitudes $G \simeq 23$ is essential to properly resolve the intrinsic kinematics of dwarf galaxies around the MW. With several thousand deployable apertures, a multiplexed spectrograph could routinely obtain radial velocities for several thousands member stars per galaxy, enabling precise reconstruction of the global velocity field and the mass distribution from the inner regions out to large radii. Such global datasets act as the dynamical “backbone”, ensuring that any central velocity anomaly is interpreted within a well-constrained gravitational framework.

In the innermost arcseconds, however, the requirements become much more stringent. The sphere of influence of IMBHs in the targeted mass regime typically spans $0.3-1$ pc, corresponding to only $\simeq0.5-1.5$ arcseconds for galaxies at distances of 80-150 kpc. Within this region, the number of available tracers is intrinsically small, often limited to a few tens of stars. An integral-field spectrograph with spatial sampling matched to natural seeing-or ideally improved by a wide - field ground-layer adaptive optics system — is therefore essential. Reaching depths equivalent to G-band magnitudes of 23 would allow secure measurements for 20-50 stars within two projected spheres of influence, depending on the target, the threshold at which dynamical models can begin to distinguish between IMBH and no-IMBH scenarios in the $\gtrsim10^4 \Msun$ range. In favorable cases, such sampling may even begin to probe the $\simeq10^3\Msun$ regime. Improving angular resolution by even a factor of two substantially increases the number of resolvable stars in the nuclear regions, reduces blending, and enhances the precision of stellar centroiding, an important consideration given that mis-centering by a few arcseconds can dilute or completely mask the IMBH signal.

The combination of a wide-field multiplexed spectrograph and a deep, high-resolution integral-field unit within the same facility is therefore not simply advantageous but indispensable. Only such a coordinated approach can provide, in a single observational framework, the large-scale dynamical context and the finely resolved central kinematics needed to identify high-velocity outliers or subtle changes in the inner velocity dispersion profile. These two components act synergistically: the wide-field data constrain the global gravitational potential, while the central spectroscopic observations probe the regime where the IMBH influence is maximal.

While this facility would stand as the primary engine of IMBH discovery, complementary observations from other platforms would play a supportive-though no foundational-role. High-resolution instruments on the ELT would offer follow-up capabilities for the most promising IMBH candidates, refining central velocity measurements or resolving crowded fields where necessary. Gaia astrometry provides membership and proper-motion information, especially for brighter stars, thereby improving the purity of the spectroscopic samples. Large spectroscopic surveys (e.g., 4MOST, WEAVE) provide global kinematic baselines but lack the depth and spatial resolution required to probe IMBH spheres of influence. In this framework, these facilities act as valuable supplements, whereas the next-generation wide-field spectroscopic facility described here is the essential component enabling a decisive exploration of the IMBH mass regime in dwarf galaxies. 

\begin{multicols}{2}
\bibliography{biblio}
\end{multicols}




\end{document}